\documentstyle[15lomcon,cite,epsfig]{article}
\include{graphics}

\input epsf
\begin{document}

\title{NUMERICAL INVESTIGATION OF LATTICE WEINBERG - SALAM MODEL}

\author{ M.A.Zubkov \footnote{e-mail: zubkov@itep.ru}}

\address{ITEP, B.Cheremushkinskaya 25, Moscow, 117259, Russia}


\maketitle\abstracts{ Lattice Weinberg - Salam model without fermions for the
value of the Weinberg angle $\theta_W \sim 30^o$, and bare fine structure
constant around $\alpha \sim \frac{1}{150}$ is investigated numerically. We
consider the value of the scalar self coupling corresponding to bare Higgs mass
around $150$ GeV. We investigate phenomena existing in the vicinity of the
phase transition between the physical Higgs phase and the unphysical symmetric
phase of the lattice model. This is the region of the phase diagram, where the
continuum physics is to be approached.  We find the indications that at the
energies above $1$ TeV nonperturbative phenomena become important in the
Weinberg - Salam model.}



The investigation of finite temperature Electroweak phase transition
\cite{EW_T} requires nonperturbative methods. The phase diagram of the lattice
Weinberg-Salam model at zero temperature also contains the phase transition.
  It is expected, that the continuum physics is
approached in some vicinity of this transition. Basing on an analogy with the
case of finite temperature Electroweak phase transition we suggest the
hypothesis that the nonperturbative effects may become important close to the
phase transition of the zero temperature model, i.e. at high enough energies
(above about $1$ TeV). This justifies the use of lattice methods in
investigation of this model. We indeed obtain some results that support the
mentioned hypothesis. Here we report these results. In our investigation we
restrict ourselves by the following bare values of couplings: fine structure
constant $\alpha \sim \frac{1}{150}$, the Weinberg angle $\theta_W = \pi/6$,
the Higgs boson mass $M_H \sim 150$ GeV. From the previous analysis \cite{BVPZ}
we know that the renormalized values of $\alpha$ and $M_H$ do not deviate
essentially from their bare values. We consider the model without fermions.  We
exclude the first order phase transition because we do not observe any sign of
a two - state signal. Also we find the indications that the second order phase
transition is present. The ultraviolet cutoff is increased when the phase
transition is approached.

\begin{figure}
\begin{center}

 \epsfig{figure=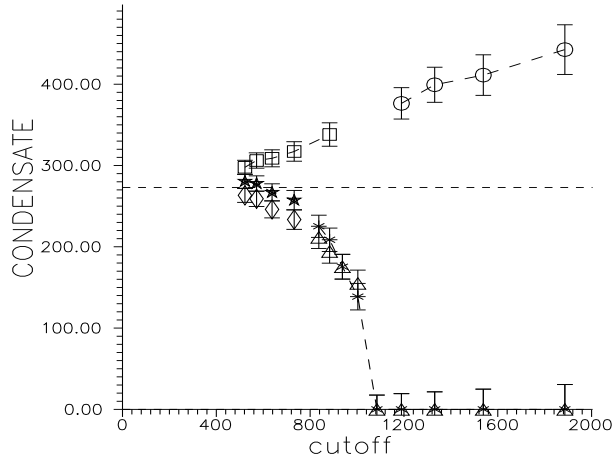,height=60mm,width=80mm,angle=0}
\caption{\label{fig.1_} The scalar field condensate (in GeV) as a function of
the cutoff
 for $\lambda =0.0025$ , $\beta = 12$.  Circles correspond to the UZ potential, lattice $16^3\times 32$.
Squares correspond to the UZ potential, lattice $8^3\times 16$. Crosses
correspond to the UDZ potential, lattice $16^3\times 32$. Stars correspond to
the UDZ potential, lattice $8^3\times 16$. Triangles correspond to the
ultraviolet potential, lattice $16^3\times 32$. Diamonds correspond to the
ultraviolet potential, lattice $8^3\times 16$. }
\end{center}
\end{figure}

We consider three different effective constraint potentials. For their
definition see \cite{BVPZ}. The three mentioned above effective potentials give
three different definitions of the scalar field condensate (as the value of
$\phi$, where the potential $V(\phi)$ has its minimum). In Fig. \ref{fig.1_} we
represent these three condensates as functions of the cutoff.

Also we calculate the percolation of the Nambu monopoles and the Z - strings.
We observe the condensation of Nambu monopoles and Z - strings for $\Lambda
> 1$ TeV and the deviation of the calculated values of the scalar field
condensates from the expected value $2M_Z/g_Z\sim 273$ GeV for the values of
$\Lambda$ around $1$ TeV. We consider these results as indications that
nonperturbative effects become important in lattice Weinberg - Salam model for
the value of the cutoff above about $1$ TeV.

This work was partly supported by RFBR grant 09-02-00338, 11-02-01227, by Grant
for Leading Scientific Schools 679.2008.2. The numerical simulations have been
performed using the facilities of Moscow Joint Supercomputer Center,
 the supercomputer center of Moscow University, and the supercomputer center of Kurchatov
 Institute.

\end{document}